\documentclass[10pt,journal,compsoc]{IEEEtran}

\usepackage{subfigure}
\usepackage{graphicx}
\usepackage{graphicx, subfigure, epsfig, multirow}
\usepackage{textcomp}
\usepackage[urlbordercolor={1 1 1} ]{hyperref}
\usepackage[hyphenbreaks]{breakurl}
\usepackage{amssymb,amsmath}
\usepackage{algorithm}
\usepackage{algpseudocode}
\usepackage{color}
\usepackage{booktabs}
\usepackage{geometry}
\usepackage{framed}

\makeatletter
\renewcommand{\ALG@beginalgorithmic}{\footnotesize}
\makeatother

\begin{document}
%
 \title{How Secure Is Your IoT Network?}
 \author{Josh~Payne,~\IEEEmembership{Student~Member,~IEEE,}
        Karan~K.~Budhraja,
        and~Ashish~Kundu,~\IEEEmembership{Senior~Member,~IEEE}
\IEEEcompsocitemizethanks{
\IEEEcompsocthanksitem Josh Payne is with Stanford University as well as the Thomas J. Watson Research Center, Yorktown Heights, NY, USA. E-mail: jfp@cs.stanford.edu\protect\\
\IEEEcompsocthanksitem Karan K. Budhraja was with University of Maryland, Baltimore County. E-mail: karanb1@umbc.edu\protect\\
\IEEEcompsocthanksitem Ashish Kundu was with the IBM Thomas J. Watson Research Center, Yorktown Heights, NY, USA. He is now with Nuro. E-mail: akundu@acm.org}%
}
\IEEEtitleabstractindextext{%
\begin{abstract}
The proliferation of IoT devices in smart homes, hospitals, and enterprise networks is widespread and continuing to increase in a superlinear manner. With this unprecedented growth, how can one assess the security of an IoT network holistically? In this article, we explore two dimensions of security assessment, using vulnerability information of IoT devices and their underlying components (\textit{compositional security scores}) and SIEM logs captured from the communications and operations of such devices in a network (\textit{dynamic activity metrics}) to propose the notion of an \textit{attack circuit}. These measures are used to evaluate the security of IoT devices and the overall IoT network, demonstrating the effectiveness of attack circuits as practical tools for computing security metrics (exploitability, impact, and risk to confidentiality, integrity, and availability) of heterogeneous networks. We propose methods for generating attack circuits with input/output pairs constructed from CVEs using natural language processing (NLP) and with weights computed using standard security scoring procedures, as well as efficient optimization methods for evaluating attack circuits. Our system provides insight into possible attack paths an adversary may utilize based on their exploitability, impact, or overall risk. We have performed experiments on IoT networks to demonstrate the efficacy of the proposed techniques.
\end{abstract}

}
\maketitle

\section{Introduction}

Statista estimates that the number of connected IoT devices will rise from 19.4B in 2018 to 34.2B in 2025 \cite{statista2019smart}, and that the percentage of U.S. homes that are \textit{smart} will rise from 33.2\% in 2019 to 53.9\% in 2023 \cite{statista2019internet}. The usage of Internet of Things (IoT) devices in networks such as smart homes, smart cities, and digital healthcare is clearly increasing, and while the adaptable and heterogeneous features of these networks have proven to be crucial in solving a vast array of different issues, they have also given adversaries a veritable sandbox of vulnerabilities to exploit. Security measures that prevent attackers from exploiting these vulnerabilities are more important now than ever before because of the growing quantity and sensitivity of data that people are putting online. The complexity of multi-stage privacy, service, and system integrity attacks is also growing, and it is now critical for home owners, enterprises, or government organizations that host various IoT devices on complex IoT networks (augmented by the increasing commonality of the Bring-Your-Own-Device (BYOD) norm) to be aware of the cybersecurity risks that may be present. In particular, it is vital to determine high-priority vulnerabilities that need to be addressed in order to keep the overall network, as well as the individual devices, protected.

We propose the use of a flow network for evaluating the security of an IoT network, which we refer to as an \textit{attack circuit}.\footnote{Code: \url{https://github.com/Josh-Payne/iot-sec-attack-circuits}} This structure arises naturally from the often modular, extensible, and heterogeneous nature of IoT networks and helps to model possible attack paths and evaluate the security state of the represented IoT network as well as each individual device therein. An attack circuit is a type of attack graph with edges defined by the inputs and outputs of its vertices, just as an electrical circuit's nodes map input and output of circuit elements. The term \textit{attack circuit} is particularly inspired because of the heterogeneity of electrical circuit elements and the ``pluggable" property of having elements added or removed. This contribution solves problems such as vulnerability documentation processing, delivering different types of metrics in the form of \textit{compositional scoring} and network flow analysis and incorporating network activity data into \textit{dynamic activity metrics}. Current methods, discussed in Section \ref{sec:related_work}, seek to address these problems for general networks as well as specifically for IoT networks---our work builds off of these ideas, adding our own novelties to address some of the limitations of current work. These novelties solve problems described in Section \ref{sec:problem_definition} such as vulnerability documentation processing, delivering different types of metrics in the form of compositional scoring and network flow analysis, and incorporating network activity data into dynamic activity metrics (this is explored further in Section \ref{sec:proposed_method}). In Section \ref{sec:implementation}, we discuss the practical and theoretical usefulness of this notion in the context of a smart home, using vulnerability data from the National Vulnerability Database (NVD)\footnote{http://nvd.nist.gov/download.cfm}, network traffic data from off-the-shelf IoT devices, and optimization and machine learning techniques for constructing and evaluating the resulting attack circuits and attack paths. Finally, in Section \ref{sec:evaluation}, we demonstrate our own implementation of the attack circuit and evaluate the effectiveness of such a notion through quantitative experimentation. This is followed by concluding remarks in Section \ref{sec:conclusion}.


\section{Related Work}
\label{sec:related_work}

In computer security, a \textit{vulnerability} is defined as a weakness of a system that can be exploited by an attacker. The attacker may then perform unauthorized activities within the system. Alternatively, an \textit{exposure} is a software error in the system that allows the attacker to gain access to system data and conduct information gathering activities. The attacker may accompany this by hiding unauthorized system activity from associated monitoring services. Subsequently, the Common Vulnerabilities and Exposures (CVE) system \cite{mell2006common} is a built reference for publicly identified information-security vulnerabilities and exposures. The system is maintained by the Mitre Corporation\footnote{mitre.org}. CVE entries are primarily composed of identifiers, descriptions, references and the date at which the CVE entry was created. The Mitre Corporation also maintains the Common Weakness Enumeration (CWE) system \cite{martin2007common}, which categorizes software weaknesses and vulnerabilities. The combined use to CVE and CWE allows organizations to select appropriate software tools for internal usage. Our work uses the CVE and CWE systems as a standardized source of information to generate a representation of possible attacks. Additionally, the Common Vulnerability Scoring System (CVSS) \cite{mell2006common} provides a way to capture the principal characteristics of a vulnerability and produce a numerical score reflecting its severity. We build on these existing scores to evaluate device and network vulnerability.

Work in \cite{apthorpe2017spying} emphasizes the security risks of smart-home networks using commercially available smart-home devices with encrypted communication. The authors explore attacks by first identifying the device (using Domain Name System (DNS) queries or device fingerprinting) and then inference of activities based on changes in network traffic. Network traffic-based threats are also highlighted in \cite{apthorpe2017smart,galtsev2011network}, where the authors demonstrate an attacker that passively observes encrypted network traffic to infer sensitive details about network users.

Work in \cite{miettinen2017iot} examines the security flaws for smart-home networks with specific interest in the exploitation of the lack of mechanisms for firmware updates or patches for security vulnerabilities. The authors propose a system to identify the types of devices that are connected and suggest the use of appropriate communication constraints, given that knowledge. The device type for their work is the enumeration of a specific device. Work in \cite{miettinen2017iot} is, however, limited to the formulation of a method to identify a given device, and does not provide a metric to determine how vulnerable a device is. Network attacks are modeled using attack graphs in \cite{ingols2009modeling}. The authors propose a scalable model zero-day exploits \cite{turner2005symantec} and client-side attacks \cite{choo2011cyber,chang2009your}, in contrast to prior attack graph systems that focus on server-side vulnerabilities \cite{noel2008optimal,lippmann2005annotated}. The work, however, focuses on modeling such attacks and corresponding countermeasures, but does not provide a means to quantify the relative impacts of different attacks. Work in \cite{matsuoka2015security} discusses the provision of security objectives for smart-home networks. While it does not discuss the underlying detection mechanism used, the application of security flaw detection is aligned with the motivation for our work.

Term Frequency-Inverse Document Frequency (TF-IDF) \cite{leskovec2014mining} is a numerical statistic used in information retrieval to determine the relative importance of words in a document. TF-IDF and TF and IDF individually may also serve as heuristics for weighting words. Our work leverages TF-IDF to compute attack meta-data for a given CVE entry. TextRank \cite{mihalcea2004textrank,PyTextRank} is a graph-based ranking model for text processing. It is inspired by recursive graph-based ranking algorithms such as HITS \cite{kleinberg1999authoritative} and PageRank \cite{page1999pagerank}, using a voting mechanism. While TextRank may be used for sentence extraction and text summarization, our work uses the information stored in the intermediate process: the extraction and ranking of phrases.

Threat, Vulnerability, Risk Analysis (TVRA) \cite{etsi2010intelligent} generates an integer value of risk based on the attack likelihood for a given network. However, TVRA encodes threats as trees, which provides less comprehensive information of interaction of attacks than methods such as flow networks explored in our work. While the Operationally Critical Threat, Asset, and Vulnerability Evaluation (OCTAVE) \cite{alberts2003introduction, alberts2005octave} framework enables risk-based strategy assessment and associated planning techniques, it requires the training of team members in order to conduct the complex analysis. Our work is aligned in purpose with the Open Web Application Security Project (OWASP) \cite{wichers2013owasp,juliadotter2015cloud} and HEAling Vulnerabilities to ENhance Software Security and Safety (HEAVENS) \cite{islam2016risk}. OWASP uses attack vectors to compute technical and business impacts to a network, producing a raw score and also an associated integer label describing attack severity. HEAVENS uses and attack probability table for threat analysis and risk assessment. Our work differs from these by using recursive information with network flow in the compositional score, potentially combined with the use of other sequential learning techniques in the dynamic activity metrics.

Additionally, attack graph generation tools are proposed by ~\cite{barrere2017naggen,sheyner2002automated}. However, they are focused on non-IoT networks. Our work is specific to IoT networks, with NLP assisted graph generation.
\label{sec:proposed_method}
\begin{figure*}[t]
    \centering
    \includegraphics[width=15cm]{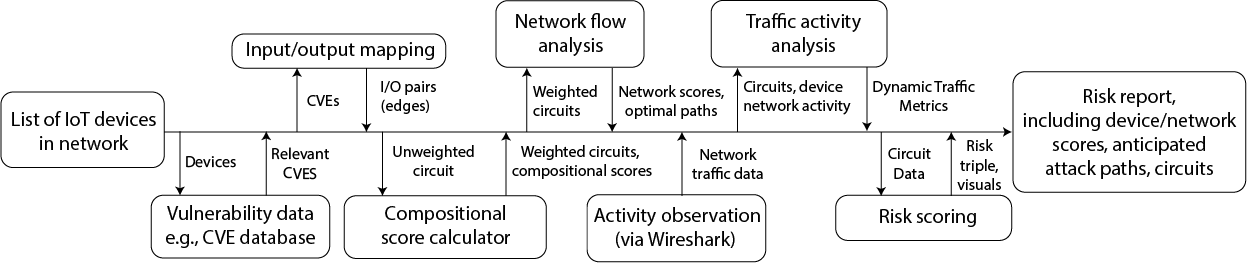}
    \caption{System architecture summary.}
    \label{fig:system_architecture}
\end{figure*}
\section{Problem Definition}
\label{sec:problem_definition}

Given a network of IoT devices and any additional knowledge about them (e.g., from CVEs or the device specification), the problem addressed by our work is to compute a security state triple $\langle R, E, I \rangle$ corresponding to the risk triple, exploitability score, and impact scores for each vulnerability, device, and the network. Exploitability is a measure of how difficult it would be for an adversary to compromise the object, and the impact is a measure of the level of harm or compromise an adversary could inflict in the case of vulnerability exploitation. Risk is meant to be interpreted as a holistic measure of the security state of the CVE, device, or network, which evaluates the confidentiality, integrity, and availability risks of the object's potential vulnerabilities: $R = \langle R_{Conf}, R_{Integ}, R_{Avail} \rangle$. $R_{Conf}$ measures the impact of a successfully exploited vulnerability on the confidentiality of information managed by the device or network. A value of \textit{Low} for $R_{Conf}$ means that there is a low risk of disclosure of such information to unauthorized individuals or systems. $R_{Integ}$ measures the impact of a successfully exploited vulnerability on the integrity of the system. For instance, if $R_{Integ}$ has a value of \texttt{Complete}, an unauthorized user may be able to easily gain root access to a device following the exploitation of an associated vulnerability. $R_{Avail}$ measures the impact of a successfully exploited vulnerability on the availability of the devices or networked services involved. This may include disk space, bandwidth/latency, and the uptime of the devices and components involved. Having a high availability risk would be particularly alarming for medical IoT networks, where lives depend on device fidelity and responsiveness.

The problem also incorporates the generation of a representation of the scored system for further assessment, including identifying possible attack paths that an adversary may traverse to carry out multi-stage attacks on the network. This may include network visuals, analysis-ready representations, lists of likely attack paths with respect to different metrics, and flow network problem solutions for downstream score computation. Such a representation would provide insights into otherwise very complex and unique IoT networks, and would give improved information for our security state triple.

Because vulnerabilities often stem from the way a system is used, the problem additionally incorporates data learned from the traffic of a particular device and holistic network behaviors. For instance, anomalous traffic volume can often be an indication of Denial-of-Service (DoS) attacks, and instances when a device is suddenly receiving responses from or sending requests to a blacklisted IP address could factor into our security state triple. This broadens the scope of our security state assessment and could provide strategies for real-time countermeasures against adversaries.










\section{Proposed Method}\label{sec:proposed_method}

Our proposal is a sequential computation of the triple $\langle R, E, I\rangle$---a method that utilizes preexisting knowledge about the devices in the network, topology of the network (determined by the relation between different vulnerabilities and potential network flow between devices), and dynamic device activity and network traffic information. An outline of the proposed system architecture is shown in Figure \ref{fig:system_architecture}. At the practical level, the attack circuit model we are proposing considers each of these properties and can provide each of the desired metrics. An attack circuit is a type of flow network \cite{goldberg1989network}, wherein the flows can be used to evaluate level of risk, exploitability, and impact of a network, device, and vulnerability. In this section, we will discuss the nontrivial problem of constructing attack circuits using preexisting knowledge and inferences about IoT devices in the network. We will then explore how the network can then be used to compute \textit{compositional scores} on each of the vulnerabilities, which provide an important baseline for giving a numerical assessment of exploitability, impact, and risk of the vulnerabilities of the network. Then, we'll examine the affect that dynamic network traffic behaviors and properties have on the security of devices in the network and show how these \textit{dynamic activity metrics} can be combined with the compositional scores for a more holistic look at the security of the IoT network. Finally, we'll propose the use of network flow algorithms that make use of the scores we've calculated for anticipating optimal attack paths, which give a final evaluation of the risk, exploitability, and impact measures of the network and its devices. 


\subsection{Attack Circuits}
An attack graph models the ways that a hacker can exploit vulnerabilities in order to carry out an attack~\cite{sheyner2002automated}. In this paper, we have proposed a notion of an \textit{attack circuit}, which is especially applicable to IoT network analysis but may also be used for other applications. An attack circuit is a class of attack graph with vertices which represent known vulnerabilities and devices, directed edges which represent the sequential exploitation possibilities of a multi-stage attack vector, edge weights which correspond to some desired security metric, and a method of evaluation that uses flow network methodologies. This gives us the first important property of attack circuits: they are \textit{metric-agnostic}; that is, they can be evaluated with respect to any metric assigned. Because circuits have inputs and outputs, it is feasible to determine if two attack circuits can be composed together to develop a more sophisticated attack circuit. In our system, we develop an attack circuit per CVE (vulnerability) and we compose the attack circuits across multiple CVEs for one or more devices on a network. This reveals the second important property of attack circuits: they are \textit{composable}; that is, one can compose $n$ attack circuits to create a single attack circuit depending on the input/output pairs described below. Attack circuits are versatile and also have applications in cloud cluster computing and other distributed systems that we will focus on in future work.

\subsection{Circuit Construction}

The attack circuit is constructed in two stages, described in the following sections.

\subsubsection{Input/Output Extraction}
\label{subsubsec:input_output_extraction}

The attack circuits are modeled using text input (vulnerability descriptions) from a vulnerability database. For each item in the database, a corresponding \textit{input/output} pair is generated. The \textit{input} corresponds to the attack source and the \textit{output} corresponds to the attack target. The process is based on TF-IDF \cite{leskovec2014mining} and TextRank \cite{mihalcea2004textrank,PyTextRank} heuristics and is described next as a series of steps.

All text is primed by conversion to lowercase, removal of non-alphanumeric characters, tokenization, stemming \cite{porter1980algorithm,jones1997readings} and subsequent de-tokenization. TF-IDF is then used on the processed corpus to produce an ordering of tokens for each description. TextRank is used on the processed corpus to produce an ordered list of candidate phrases (with Part-Of-Speech (POS) tags) that may best represent the description. The ordered list is filtered to remove noun items (NN). For each item in the list, tokens are pruned (limited to a maximum quantity of $3$) based on TF-IDF ordering. Stemming is then removed from tokens by matching them to the corresponding phrase. The result is stored as \textit{input}. This extraction process is then repeated using filtering to remove non-noun items and the result is stored as \textit{output}.

\subsubsection{Graph Composition}

After all of the \textit{input/output} pairs are created, we have the information we need to build the attack circuit structure. In our methodology, an attack circuit is a directed graph isomorphic to a flow network $C = (D,A,S,E)$ where $d \in D$ is a device, represented as a set of vertices that are the set of vulnerability database entries corresponding to $d$; $A$ is the set of attacker vertices; $S$ is the set of target (or sink, as we'll see later) vertices which represent the attack targets in the network; and $E$ is the set of labeled, directed edges. These edges are weighted with the associated vulnerabilities' impact and exploitability base scores. The attack circuit scheme suffices to provide a logical approach to determining which attack targets are at risk. We are also interested in creating variants of the attack circuit that may give insight to the potential impact, exploitability, and overall risk an IoT network yields.

\subsection{Network Composition Analysis}

To take a holistic view of the system, we use several different security metrics that may each be classified into one of two larger categories: compositional scoring and dynamic activity assessment, discussed in the next subsection. Compositional scoring may be performed by observing the device specification, associated vulnerability database information and their corresponding vulnerability scores, and attack circuit topology. Our work incorporates the first two items and designates the third for future work. Compositional scores are derived from the devices and their role in the network composition, and can be calculated irrespective of ways they are used over time. Vulnerabilities are scored using the base risk, exploitability, and impact subscore method provided by the existing CVSS v3 standard. After computing the exploitability and impact subscores for each vulnerability, we can then calculate the compositional score ($c$), which is recursively computed using the impact weights and exploitability capacities of the edges of the graph. This is summarized in Equation \ref{eq:exploitability_score} and Equation \ref{eq:impact_score}, where $C_i$ is the set of attack circuit input vertices ($c_i$), where there exists an edge $(c_i,c)$, $C_o$ is the set of attack circuit output vertices ($c_o$), where there exists an edge $(c,c_o)$, and $v_d$ is a dampening constant. The impact and exploitability subscores of each of the device's vulnerabilities are then used to determine the impact and exploitability subscores.

\begin{equation}
c_{Exploitability} \mathrel{+}= v_d\sum_{c_i}c_{i_{Exploitability}}
\label{eq:exploitability_score}
\end{equation}

\begin{equation}
c_{Impact} \mathrel{+}= v_d\sum_{c_o}c_{o_{Impact}}
\label{eq:impact_score}
\end{equation}

\subsection{Network Traffic Analysis}

We next examine the dynamic activity metric of a given device in the network. A large body of work is focused on abnormality detection and scoring in network traffic patterns \cite{acar2018peek,apthorpe2017spying,apthorpe2017smart}. A vast variety of metrics may be used for results that shed light onto particular aspects of IoT device and network security. Our work focuses on ascertaining a small number of these metrics from packet sniffing on the network of IoT devices over a period time. The data collected includes device traffic activity, packet content encryption, and the source/destination information of the packets.

\begin{figure}[t]
    \centering
    \includegraphics[width=0.48\textwidth]{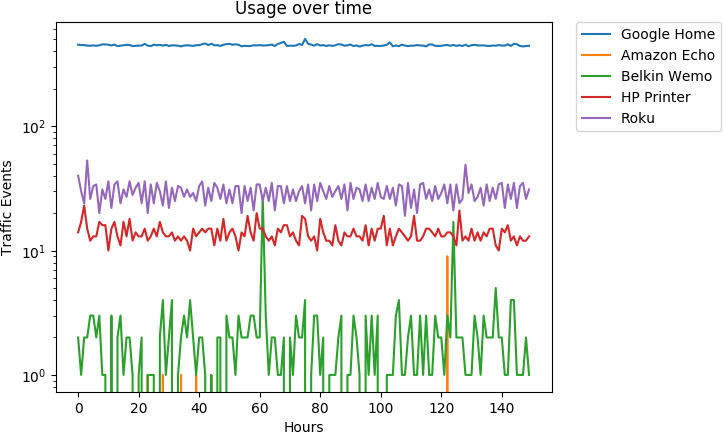}
    \caption{An account of device traffic over a period of $12$ hours.}
    \label{fig:usage_over_time}
\end{figure}

To illustrate our use of device traffic activity information, consider the example shown in Figure \ref{fig:usage_over_time}. Devices that are online for the majority of the time (i.e. Google Home Mini, Roku Media Player, HP Printer, and Belkin WeMo) are assigned larger respective multipliers to their exploitability subscore, since their connection leaves them more open to attack by an adversary. The Amazon Echo Dot, in this case, does not have as significant a multiplier applied to its exploitability subscore, since it is not online as frequently.

Next, we analyze the percentage of the packets that are sent from and received by each device and their usage of secure encryption protocols. We also check whether any source or destination IP is listed in an IP blacklist database. The encryption metric is used as a multiplier for the exploitability subscore, and the blacklisted IP metric is used as a multiplier for the impact subscore. The calculated compositional scores and dynamic activity metrics for a device can then be used to improve security risk scoring and attack path analysis.

\subsection{Network Flows and Attack Path Analysis}\label{nwflows}

To anticipate how an adversary might carry out an attack on the network and to score the network holistically, we apply a variety of \textit{Network Flow Problems} \cite{ford2009maximal} \cite{dantzig2003max} to the attack circuit for evaluating potential attack paths based on impact, exploitability, and risk. Sources and sinks in the flow network correspond to attack sources and attack targets, respectively.

\subsubsection{Impact Paths} 

An impact path is the route through the circuit that an attacker takes to maximize impact, defined by the circuit edge weights. We specify the attacker nodes as sources, and attacker targets as sinks. Equation \ref{eq:impact_path} illustrates the \textit{Maximum Flow Problem} method used---$f_{uv}$ is the flow between vertices $u$ and $v$, $a$ is an attacker vertex, $s$ is a sink (target) vertex. The sum of the impact path flows of an attack circuit (or the impact score of that circuit) are equivalent to the sum of the total impact of all \textit{easily} accessible attack targets, where the degree of accessibility is defined by the exploitability of the attack paths leading to it.

\begin{equation}
\begin{array}{ll@{}ll}
\text{maximize} & \displaystyle\sum\limits_{} f_{as} & \text{ subject to } \displaystyle\sum\limits_{j} f_{ji} = \displaystyle\sum\limits_{j} f_{ij}
\end{array}
\label{eq:impact_path}
\end{equation}

\subsubsection{Exploitability Paths} 

An exploitability path is a route from attacker to the attack target that is associated with a score  denoting the \textit{resistance} (intuitively, inverse exploitability) of that path. To determine the optimal exploitability paths in a circuit, we solve a \textit{Minimum Cost Flow Problem}, where the cost of an edge is its resistance, or the inverse of the exploitability of that edge (e.g. ($1-$Exploitability), if Exploitability $\in [0,1]$). We use the same sources and sinks as in Equation \ref{eq:impact_path} now with a cost $c_{ij}$ associated with edge $(i,j)$ and a required flow $r_{as}$ from attacker to sink (see Equation \ref{eq:exploitability_path}).

\begin{equation}
\begin{array}{ll@{}ll}
\text{minimize}  & \displaystyle\sum  c_{ij}f_{ij}
\text{ subject to } & \displaystyle\sum f_{as} = r_{as}
\end{array}
\label{eq:exploitability_path}
\end{equation}\\
After computing the optimal exploitability and impact paths, the paths themselves may serve as information for network operators. We may also sum over the exploitability paths in the network to determine an overall network exploitability score, which we then combine with the impact score to improve an overall network security risk score.

\subsubsection{Risk Flow Evaluation} 
Finally, we want to calculate high-risk paths and quantitatively apply these findings to our risk triple. A risk path is a route from attacker to target that is optimized for exploitability constraints, impact weights, and base compositional risk of the vertices. We combine the strategies used in exploitability path and impact path analysis here, solving a \textit{Minimum Cost Maximum Flow Problem} to identify likely paths through the IoT network that an adversary might use in an attack. In Equation \ref{eq:risk_mincost_path}, we outline the network flow problem used to calculate risk, with the usual $c_{ij}$ denoting cost of edge $(i,j)$, $a$ being an attacker vertex (source), and $s$ being a target vertex (sink):
\begin{equation}
\begin{array}{ll@{}ll}
\text{minimize } \displaystyle\sum c_{ij}f_{ij}
\text{ subject to: } \\
\ \ \ \ \text{maximize } \displaystyle\sum\limits_{} f_{as} \text{ subject to } \displaystyle\sum\limits_{j} f_{ji} = \displaystyle\sum\limits_{j} f_{ij}
\end{array}
\label{eq:risk_mincost_path}
\end{equation}
Computing risk paths allow us to calculate an improved risk triple for the network as well as the devices within. The triple $R = \langle R_{Conf}, R_{Integ}, R_{Avail} \rangle$ for a device is calculated using the CVSS v3 metrics for Confidentiality Impact, Integrity Impact, and Availability Impact associated with the CVEs of the device. Flow through each CVE is multiplied by the respective impact metric, and the result defines the risk triple $R$. The numerical values of the impact metrics may be found in table 8.4 of the CVSS v3 specification document\footnote{https://www.first.org/cvss/specification-document}.
    
  

\section{Implementation}
\label{sec:implementation}

\subsection{Our System}

\begin{figure}[t]
    \centering
    \includegraphics[width=0.48\textwidth]{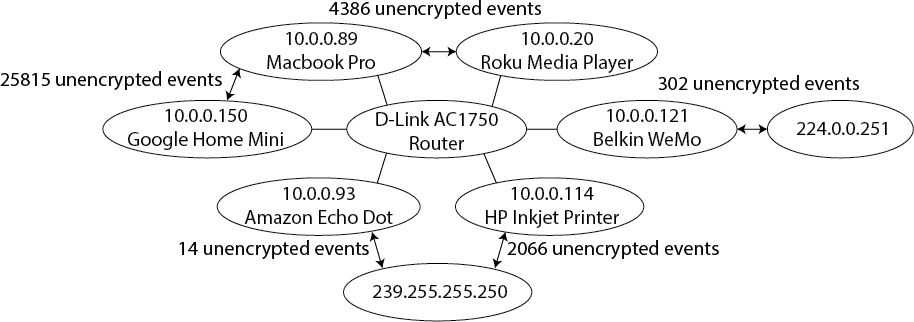}
    \caption{The smart-home network traffic graph corresponding to Figure \ref{fig:usage_over_time}.}
    \label{fig:smart_home_network}
\end{figure}

To understand the performance of our proposed method in common, real-world networks, we modeled our implementation with off-the-shelf smart home IoT devices, narrowing our focus to the $34$ devices that are common devices and also have corresponding vulnerabilities listed in the NVD. Live experiments were run on a network consisting of five of these devices: an Amazon Echo Dot, a Belkin WeMo, an HP Inkjet Envy printer, a Google Home Mini, and a Roku digital media player (see Figure \ref{fig:smart_home_network}).

Circuit construction is implemented using relevant CVEs in JSON format from the NVD, with text processing using the Natural Language Toolkit (NLTK) \cite{bird2004nltk}. An example of the CVE processed as in Section \ref{subsubsec:input_output_extraction} is shown in Figure \ref{fig:processed_cve}. In this example, an \textit{input} (left of the arrow in the \texttt{i/o} field) is an action (DNS rebinding attack) that the attacker needs to take to leverage the corresponding vulnerability, in this case, CVE-2018-11314 (for access to root privileges or the configuration file). For each input, there is an \textit{output} (right of the arrow in the \texttt{i/o} field), which indicates the target that the attacker receives when once the input is applied to the vulnerability. A device may have multiple corresponding CVEs, a CVE may have multiple corresponding inputs, and an input may have multiple corresponding outputs. NetworkX \cite{hagberg2008exploring} and SNAP \cite{leskovec2016snap} are then used to build the attack circuit using these \textit{input/output} pairs and proceed with scoring.

\begin{figure}[t]
    \centering
    \begin{minipage}{0.48\textwidth}
    \begingroup\ttfamily\obeylines
    "Roku Media Player": [\{
    \ "description": "The External Control API in Roku and Roku TV products allow unauthorized access via a DNS Rebind attack.", 
    \ "id": "CVE-2018-11314", 
    \ "i/o": ["DNS Rebinding->this:Root 
    \ Priv", "DNS Rebinding->this:Config 
    \ File"]  \}]
    \endgroup
    \end{minipage}
    \caption{An example of a processed CVE data entry.}
    \label{fig:processed_cve}
\end{figure}


\subsection{Our Scoring Method}

Let $EB_c \in [0,10]$ denote CVE $c$'s base exploitability score and $IB_c \in [0,10]$ denote CVE $c$'s base impact score. We use impact and exploitability scoring guidelines set by the NVD in their CVSS v3 method\footnote{http://nvd.nist.gov/cvss.cfm}. The numerical possibilities for the CVSS metrics $AV$, $AC$, $PR$, $UI$, $I_{Conf}$, $I_{Integ}$, and $I_{Avail}$ can be found in Table 8.4 of the CVSS specification document. Scoring during the circuit construction phase is computed as shown in Equation \ref{eq:scoring_circuit_construction}.

\begin{equation}
\begin{array}{ll@{}ll}
EB_c = 8.22 \times AV \times AC \times PR \times UI \\
ISC_{Base} = 1 - [(1-I_{Conf}) \times (1-I_{Integ}) \times (1-I_{Avail})] \\
\texttt{if Scope = unchanged}: IB_c = 6.42 \times ISC_{Base} \\
\texttt{else}: \ \ \  IB_c = 7.52 \times [ISC_{Base}-0.029] \\
\ \ \ \ \ \ \ \ \ \ \ \ \ \ \ \ \ \ \ \ - 3.25 \times [ISC_{Base}-0.02]^{15}
\end{array}
\label{eq:scoring_circuit_construction}
\end{equation}
Here, \texttt{Scope} refers to the ability for a vulnerability in one component or device to impact resources beyond its privileges. Each CVE contains a \texttt{Scope} field with a value of \texttt{changed} (if the exploited vulnerability can affect resources beyond its authorized privileges) or \texttt{unchanged} (if the vulnerable component is the same as the impacted component).

Now, let $EC_d$ denote device $d$'s compositional exploitability score and $IC_d$ denote device $d$'s compositional impact score. Let $c_{inputs}$ denote the set of CVEs with input to $c$ (where each device $d$ has a set of $c$ corresponding to each CVE), let $c_{outputs}$ denote the set of CVEs that $c$ has output to, let $f_{ij}$ be retreived by solving the appropriate network flow problem described in \ref{nwflows}, and let $v_d$ denote a dampener variable (we used $v_d$ = 0.1). Scoring during the compositional phase is computed as shown in Equation \ref{eq:scoring_compositional}.

\begin{equation}
\begin{array}{ll@{}ll}
EC_d = {\displaystyle\sum_{c \in d}(EB_c+v_d\displaystyle\sum_{c_i\in c_{inputs}}{f_{c_ic}EB_i}}) \\
IC_d = {\displaystyle\sum_{c \in d}(IB_c+v_d\displaystyle\sum_{c_o\in c_{outputs}}{f_{cc_o}IB_o}}) \\
\end{array}
\label{eq:scoring_compositional}
\end{equation}

Next we compute the network traffic multipliers for impact and exploitability. Network traffic data is collected using Wireshark \cite{orebaugh2006wireshark} and is stored in .pcap format. This file is then converted to CSV for parsing. The data is specific to a network of $5$ IoT devices and a period of $4$ days. We use three metrics: device Network Uptime ($NU$), Encryption Scheme ($EN$), and whether IP sources or destinations were listed in an IP blacklist database\footnote{https://myip.ms/browse/blacklist} ($IP$). The weights we assigned to the different categories correspond to importance with respect to the score in consideration---for instance, when considering the exploitability score, the Network Uptime multiplier = $NU$ = \texttt{\{"always\_online": 1.6, "frequently\_online": 1.4, "rarely\_online": 1.07, "never\_online": 1\}}.


The final exploitability and impact scores $E_d$, $I_d$ for a device $d$ are then calculated as shown in Equation \ref{eq:final_scores}, where $m_{E_i}$ is an exploitability-related network traffic multiplier (e.g. $NU$, $EN$), $m_{I_j}$ is an impact-related network traffic multiplier (e.g. $IP$), and $v_{n}$ is a normalizing variable (we use $v_{n_1}=v_{n_2}=100$). These scores are normalized by some sigmoidal function defined on the non-negative reals: $\sigma : \mathbb{R}_{\geq 0} \rightarrow [0,1)$.
\begin{equation}
\begin{array}{ll@{}ll}
E_d = \sigma{(v_{n_1}\times EC_d \times \displaystyle\prod_i m_{E_i})} \\
I_d = \sigma{(v_{n_2}\times IC_d \times \displaystyle\prod_j m_{I_j})} \\
\end{array}
\label{eq:final_scores}
\end{equation}

The overall network exploitability score $E_N$ is defined by the sum of the exploitability scores of its devices, and is accompanied with the path of minimum cost to each of the attack targets. The overall network impact score $I_N$ is the solution to the max-flow problem in the attack circuit after each of the edges have been weighted based on all of the CVEs' base impact scores. Finally, the risk triple $R_N = \langle R_{Conf}, R_{Integ}, R_{Avail} \rangle$ for the network is computed using the CVSS v3 metrics associated with the respective risk triples of the network's devices. First we solve the Maximum Flow Minimum Cost Problem in the attack circuit (see Equation \ref{eq:risk_path_1}). 

\begin{equation}
\begin{array}{ll@{}ll}
\text{minimize } \displaystyle\sum EB_{ij}f_{ij}
\text{ subject to: } \\
\ \ \ \ \text{maximize } \displaystyle\sum\limits_{} IC_{as} \text{ subject to } \displaystyle\sum\limits_{j} f_{ji} = \displaystyle\sum\limits_{j} f_{ij}
\end{array}
\label{eq:risk_path_1}
\end{equation}

Once this each $f_{ij}$ is computed, the flow through each CVE is multiplied by the respective impact metric---Confidentiality Impact ($I_{Conf_{c}}$), Integrity Impact ($I_{Integ_{c}}$), and Availability Impact ($I_{Avail_{c}}$) for vulnerability $c$ given by CVSS v3, as well as a normalization variable $v_{n_i}$, and the result defines the risk triple $R_d$ for a device $d$ with as set of CVEs $c_d$, each of which have a set of neighbors $c_{inputs}$ (shown in Equation \ref{eq:risk_path_2}). These scores are normalized to range $[0,1)$.

\begin{equation}
\begin{array}{ll@{}ll}
R_{Conf} = \sigma{\left(v_{n_3}\times \displaystyle\sum_{c\in d} I_{Conf_{c}} \left(\displaystyle\sum_{c_i\in c_{inputs}} f_{c_ic}\right)\right)} \\
R_{Integ} = \sigma{\left(v_{n_4}\times \displaystyle\sum_{c\in d} I_{Integ_{c}} \left(\displaystyle\sum_{c_i\in c_{inputs}} f_{c_ic}\right)\right)} \\
R_{Avail} = \sigma{\left(v_{n_5}\times \displaystyle\sum_{c\in d} I_{Avail_{c}} \left(\displaystyle\sum_{c_i\in c_{inputs}} f_{c_ic}\right)\right)} \\
\end{array}
\label{eq:risk_path_2}
\end{equation}
\section{Evaluation}
\label{sec:evaluation}
\begin{table}[t]
    \centering
    \begin{tabular}{| c | c | c | c |}
    \toprule
    {} & Echo, 1 CVE & Echo, all CVEs & Echo, WeMo \\
    \midrule
    $E_{Echo}$ & 0.0289 & 0.1182 & 0.3380 \\
    $I_{Echo}$ & 0.0140 & 0.0679 & 0.1776 \\
    Echo $R_{Conf}$ & 0.0073 & 0.0341 & 0.0982 \\
    Echo $R_{Integ}$ & 0.0 & 0.0268 & 0.0910 \\
    Echo $R_{Avail}$ & 0.0 & 0.0 & 0.0644 \\
    \midrule
    $E_{WeMo}$ & N/A & N/A & 0.8490 \\
    $I_{WeMo}$ & N/A & N/A & 0.4823 \\
    WeMo $R_{Conf}$ & N/A & N/A & 0.5744 \\
    WeMo $R_{Integ}$ & N/A & N/A & 0.5649 \\
    WeMo $R_{Avail}$ & N/A & N/A & 0.4605 \\
    \midrule
    $E_{Network}$ & 0.0289 & 0.1182 & 0.9223 \\
    $I_{Network}$ & 0.0140 & 0.0679 & 0.6078 \\
    Network $R_{Conf}$ & 0.0073 & 0.0341 & 0.6367 \\
    Network $R_{Integ}$ & 0.0 & 0.0268 & 0.6239 \\
    Network $R_{Avail}$ & 0.0 & 0.0 & 0.5098 \\
    \bottomrule
    \end{tabular}
    \\
    \caption{Device and network scores for different network settings.}
    \label{tab:results}
\end{table}
\begin{figure}[t]
    \centering
    \includegraphics[width=0.35\textwidth]{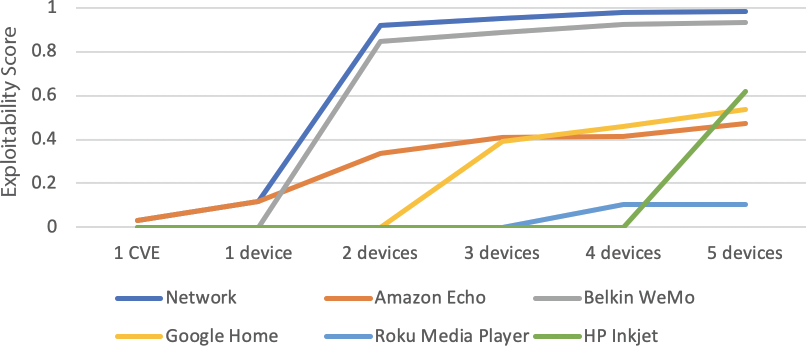}
    \caption{Exploitability score adjustment with the addition of new devices (augmented in sequence (left to right, top to bottom)) and CVEs.}
    \label{fig:exploitability_trends}
\end{figure}
\begin{figure}[t]
    \centering
    \includegraphics[width=0.20\textwidth]{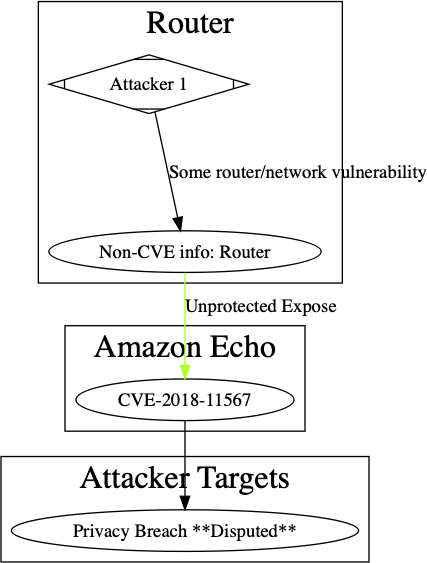}
    \includegraphics[width=0.20\textwidth]{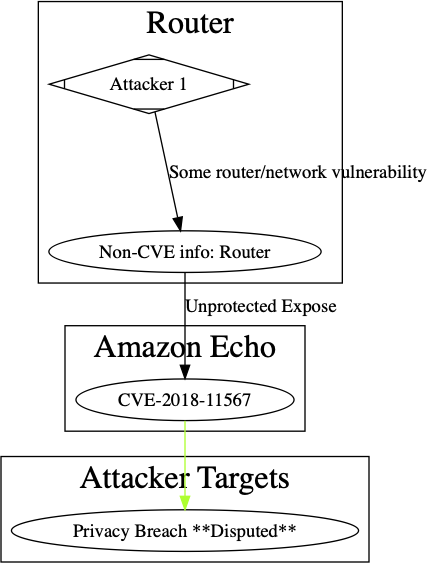}
    \caption{Exploitability (left) and impact (right) circuits corresponding to one device under the observation of a single vulnerability. For all of the circuits listed, colors correspond to score: green is the lowest, followed by yellow, orange, red, and purple being the highest. Black edges do not have an associated score.}
    \label{fig:exp1cve}
\end{figure}


\begin{figure*}[t]
    \centering
    \includegraphics[width=12cm]{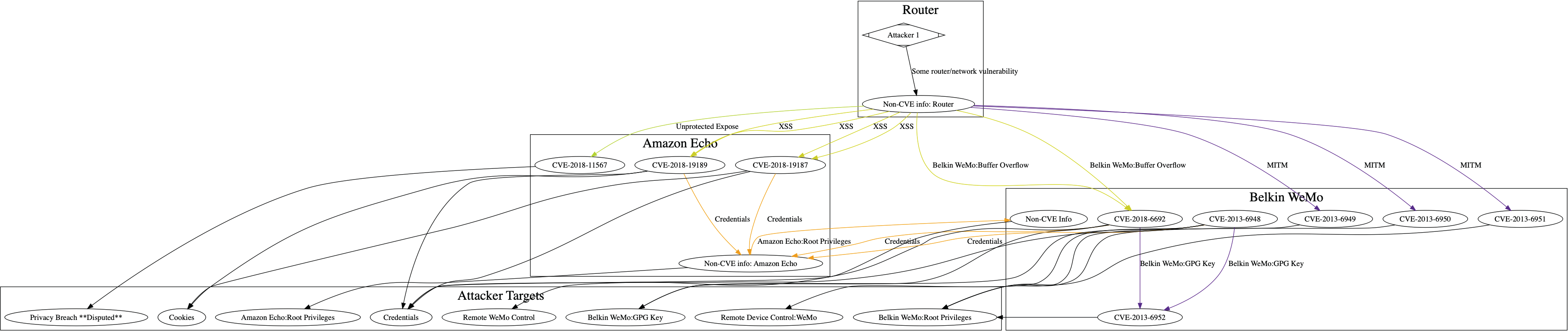}
    \caption{Exploitability circuit corresponding to two devices under the observation of all vulnerabilities. Full-sized images are available in the repository.}
    \label{fig:exp2dev}
\end{figure*}

We evaluate the attack circuit system using a variety of networks and device activity metrics. In our work, we showcase three of these networks: one consisting of one device (an Amazon Echo Dot) with one vulnerability (CVE-2018-11567), one consisting of the Amazon Echo Dot and all of its corresponding CVEs, and one consisting of the Amazon Echo Dot, a Belkin WeMo smart plug, and all of their associated CVEs. The generated exploitability and impact circuits for each of the three settings are shown at the back of this paper in Figures \ref{fig:exp1cve}, and an exploitability circuit is shown for the two-device configuration \ref{fig:exp2dev}. Note that the attack circuit complexity grows exponentially with respect to the number of devices in the network, which is visually evident in these figures and is the reason why we primarily focus on smaller networks to demonstrate our work. However, we ran a runtime test of our system on $146$ IoT device CVEs (more than most smart homes would have) and this calculation took less than a second on a laptop computer. This is not to demonstrate a primary strength of the work, but rather to point to the computational practicality of the system.

In Table \ref{tab:results}, we record the scores of devices and the network as they were calculated with each of the networks. These results reflect the dynamic activity metrics that we recorded in our experimental observation of the devices over a period of $4$ days, and the output values respond accordingly as we experimentally change the dynamic metrics of the devices (for instance, in contrast to the 1-CVE, 1-device network yielding an exploitability score of $0.0289$ with the Echo's actual status of ``\texttt{rarely\_online}", we noted that the exploitability score rose to $0.0378$ when the device was marked as ``\texttt{frequently\_online}"). Note that where only one device is used, the Belkin WeMo scores are N/A because the device is not present in that network. We select these particular devices to evaluate and illustrate the change in scores when two devices of different levels of vulnerability are added to a network. Based on its three total CVEs, the Amazon Echo's vulnerabilities don't yield as high of a risk, exploitability, or impact score of the device. Note that in the first column, we observe a network with a vulnerability that is of relatively lesser concern (CVE-2018-11567, with a base exploitability score of $1.8$ and base impact score of $1.4$). When more vulnerabilities are added in, the device's scores all increase as expected, and the network score increases in the same way (because in this setting, the network is defined solely by the device). When the WeMo is added, several paths between the WeMo's vulnerabilities and the Echo's vulnerabilities are discovered from the I/O mapping step, which causes each of the Echo's scores to rise, and the network's scores likewise undergo an increase.

This then raises the following question: how can the scores of networks, which are largely heterogeneous and unique, be interpreted relativistically? For now, we normalize the score to a range of $[0,1)$ using a sigmoidal function. In general, as the number of devices grows and the vulnerabilities increase, the score trends will demonstrate a sigmoidal behavior, converging to $1$. In Figure \ref{fig:exploitability_trends}, we observe one metric---the exploitability score---of $5$ devices and the overall network, and how it changes as more devices are added. In particular, we start with the Amazon Echo with one vulnerability, then add the rest of the Amazon Echo's vulnerabilities, then the Belkin WeMo, Google Home Mini, Roku media player, and finally HP Inkjet printer in sequence. As expected, the network's exploitability score approaches a value of $1$ as more devices are added. The case may be made that this scoring method is too sensitive (regardless of the fact that these networks are comprised of devices whose vulnerabilities we know). This may be a result of the dampener and normalization variable values or lack of information about attack circuit behavior in practice. We leave the related refinement of the scoring method to future work.

\section{Conclusion}
\label{sec:conclusion}

In this paper, we address the problem of evaluating the security of a network. This is done by using \textit{attack circuits} and associated \textit{compositional scores} and \textit{dynamic activity metrics}. In this manner, an individual IoT device or network may be analyzed for its vulnerability to security attacks. Evaluation in Section \ref{sec:evaluation} demonstrates the increased security risks for a growing IoT network. While our work focuses on using descriptions to extract \textit{input/output} pairs, this approach may be extended to extract multiple pairs per description, as well as using other available information sources. Activity metrics may be further developed by using generative machine learning models to learn abnormal network traffic behaviors. We also noted that as networks grow, their complexity grows exponentially. Thus, the network flow problems may become inefficient with huge IoT attack circuits, and other approaches such as graph neural networks may be required for analysis. However, for application in smaller IoT networks (e.g. smart homes), we conclude that this approach suffices. Further, the concept of an attack circuit is widely adaptable and may be tailored to specific use-cases in IoT networks and beyond.



\bibliographystyle{IEEEtran}
\bibliography{IEEEabrv,references}
\begin{IEEEbiography}[{\includegraphics[width=1in,height=1.25in,clip,keepaspectratio]{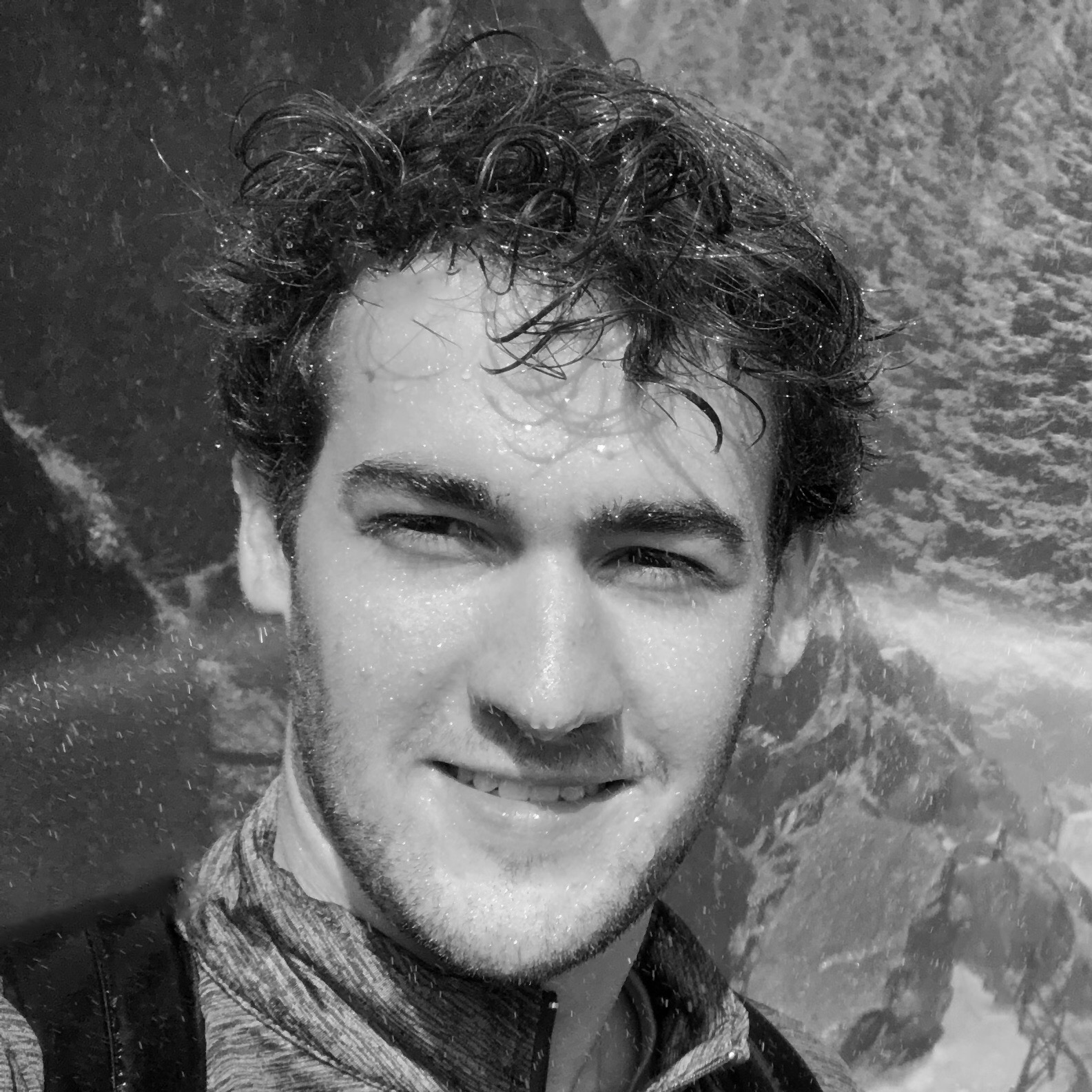}}]{Josh Payne} is currently at Stanford University, where he is advised by Bradley Efron and Dan Boneh. He is also affiliated with the IBM Thomas J. Watson Research Center, where he was recently awarded an IBM Inventor Plateau award for filing patents in identity fraud detection and digital identity management. Here, he's currently focused on learning techniques on graphs and hypergraphs, as well as differential and inferable privacy. His scientific interests include computer security, quantum computing, and deep learning on graphs and hypergraphs. He is an IEEE Student Member.
\end{IEEEbiography}
\begin{IEEEbiography}[{\includegraphics[width=1in,height=1.25in,clip,keepaspectratio]{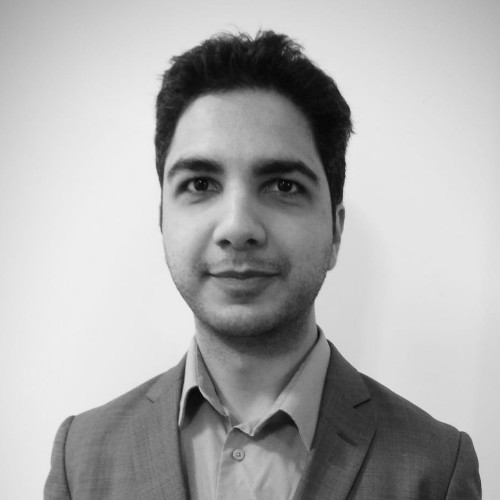}}]{Karan K. Budhraja} is a graduate from the University of Maryland, Baltimore County (receiving his M.S. and Ph.D. degrees in 2016 and 2019 respectively). His scientific interests broadly include artificial intelligence and machine learning, with recent interest in solving data security and privacy problems. His dissertation focuses on replicating emergent behaviors in agent-based models using abstract demonstrations. While the primary objective focuses on low data scenarios, a scalable neural network parallel is also devised. He is currently pursuing a career as a Data Scientist.
\end{IEEEbiography}
\begin{IEEEbiography}[{\includegraphics[width=1in,height=1.25in,clip,keepaspectratio]{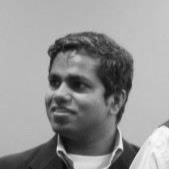}}]{Ashish Kundu} is currently Head of Cybersecurity at Nuro.AI. He is an ACM Distinguished Member and ACM Distinguished Speaker. Prior to joining Nuro, Ashish was a Master Inventor and Research Staff Member in Security Research at IBM Thomas J. Watson Research Center, Yorktown Heights, New York. Dr. Kundu received Ph.D. in Computer Science from Purdue University in 2010. His work has led to over 120 patents filed with over 100 patents granted, and over 40 research papers. He was recognized with the prestigious CERIAS Diamond Award for his outstanding contribution to cybersecurity research during his Ph.D. He is an IEEE Senior Member.
\end{IEEEbiography}
\end{document}